\documentclass[conference]{IEEEtran}
\IEEEoverridecommandlockouts

\usepackage{cite}
\usepackage{amsmath,amssymb,amsfonts}
\usepackage{algorithmic}
\usepackage{graphicx}
\usepackage{textcomp}
\usepackage{xcolor}
\def\BibTeX{{\rm B\kern-.05em{\sc i\kern-.025em b}\kern-.08em
    T\kern-.1667em\lower.7ex\hbox{E}\kern-.125emX}}

\usepackage{glossaries}
\usepackage{cleveref}
\usepackage{bm} 
\usepackage{bbm}

\DeclareMathOperator*{\Ex}{\mathbb{E}}

\DeclareMathOperator{\one}{\mathbbm{1}}

\newacronym[plural=DSOs,firstplural=Distribution System Operators (DSOs)]{dso}{DSO}{Distribution System Operator}
\newacronym{andes}{ANDES}{Advanced Net DEcision Support}

\newacronym{irena}{IRENA}{International Renewable Energy Agency}

\newacronym{hl1}{HLI}{Hierarchical Level I}
\newacronym{hl2}{HLII}{Hierarchical Level II}
\newacronym{hl3}{HLIII}{Hierarchical Level III}

\newacronym{admd}{ADMD}{After Diversity Maximum Demand}
\newacronym[plural=OSs,firstplural=\textit{Onderstations} (OSs)]{os}{OS}{\textit{Onderstation}}
\newacronym[plural=MSRs,firstplural=\textit{Middenspanningsruimtes} (MSRs)]{msr}{MSR}{\textit{Middenspanningsruimte}}
\newacronym[plural=MS-HLDen,firstplural=\textit{Middenspannings-Hoofdleidingen} (MS-HLDen)]{mshld}{MS-HLD}{\textit{Middenspannings-Hoofdleiding}}
\newacronym[plural=LS-HLDen,firstplural=\textit{Laagspannings-Hoofdleidingen} (LS-HLDen)]{lshld}{LS-HLD}{\textit{Laagspannings-Hoofdleiding}}
\newacronym{lv}{LV}{Low-Voltage}
\newacronym{mv}{MV}{Medium-Voltage}
\newacronym{hv}{HV}{High-Voltage}

\newacronym{cmc}{CMC}{Crude Monte Carlo}
\newacronym{is}{IS}{Importance Sampling}
\newacronym{ce}{CE}{Cross-Entropy}

\newacronym{evt}{EVT}{Extreme Value Theory}
\newacronym{bm}{BM}{Block Maxima}
\newacronym{pot}{POT}{Peak Over Threshold}
\newacronym{gev}{GEV}{Generalised Extreme Value}
\newacronym{gpd}{GPD}{Generalised Pareto Distribution}

\newacronym{kvk}{KvK}{Netherlands Chamber of Commerce, \textit{Kamer van Koophandel}}
\newacronym{sjv}{SJV}{\textit{Standaard Jaarverbruik}}
\newacronym[plural=EVs,firstplural=Electric Vehicles (EVs)]{ev}{EV}{Electric Vehicle}
\newacronym[plural=HPs,firstplural=Heat Pumps (HPs)]{hp}{HP}{Heat Pump}
\newacronym{pv}{PV}{Solar Photovoltaic}
\newacronym{cbs}{CBS}{\textit{Centraal Bureau voor de Statistiek}}
\newacronym{ecn}{ECN}{\textit{Energieonderzoek Centrum Nederland}}
\newacronym{nedu}{NEDU}{\textit{Nederlandse Energiedatauitwisseling}}

\newacronym{iid}{i.i.d.}{independent and identically distributed}
\newacronym{mc}{MC}{Monte Carlo}
\newacronym[plural=PMFs,firstplural=Probability Mass Functions (PMFs)]{pmf}{PMF}{Probability Mass Function}
\newacronym[plural=PDFs, firstplural=Probability Density Functions (PDFs)]{pdf}{PDF}{Probability Density Function}

\newacronym{face}{FACE}{Fully Automated Cross-Entropy}

\newglossaryentry{angelsperarea}{
  name = $a$ ,
  description = The number of angels per unit area,
}
\newglossaryentry{numofangels}{
  name = $N$ ,
  description = The number of angels per needle point
}
\newglossaryentry{areaofneedle}{
  name = $A$ ,
  description = The area of the needle point
}

\newglossaryentry{p_crit}{
  name = $p_{crit}$ ,
  description = The critical power value of each grid asset which should not be exceeded. It is defined as the power rating of an asset plus a percentage margin.
}

\newglossaryentry{p_cap}{
  name = $p_{cap}$ ,
  description = The power rating of an asset.
}

\newglossaryentry{Y}{
  name = $Y$ ,
  description = A generic random variable denoting a model output
}

\newglossaryentry{r}{
  name = $r$ ,
  description = A generic risk metric of the form \ensuremath{r = \Ex [M(\bm{X})]}
}

\newglossaryentry{r_hat}{
  name = \ensuremath{\hat{r}} ,
  description = An estimate of \ensuremath{r}
}

\newglossaryentry{R_hat}{
  name = \ensuremath{\hat{R}} ,
  description = {An estimator of \ensuremath{r}, describing the procedure to obtain an estimate}
}

\newglossaryentry{norm}{
  name = \ensuremath{\mathcal{N}(\mu, \sigma^2)},
  description = The normal distribution with expectation \ensuremath{\mu} and standard deviation \ensuremath{\sigma}
}

\newglossaryentry{S_hat}{
  name = \ensuremath{\hat{S}^2} ,
  description = The estimator of the sample variance
}

\newglossaryentry{SE}{
  name = $SE$ ,
  description = The standard error of an estimator or an estimate
}

\newglossaryentry{RE}{
  name = $RE$ ,
  description = The relative error or coefficient of variation of an estimator or an estimate
}

\newglossaryentry{Omega}{
  name = $\Omega$ ,
  description = {A sample space. In most cases, the sample space of the demand model referred to}
}

\newglossaryentry{omega}{
  name = $\omega$ ,
  description = A particular state or realisation of a sample space
}

\newglossaryentry{Theta}{
  name = $\Theta_j$ ,
  description = {The intermediate region of interest in the sample space of the $j$-th iteration. Therefore, \ensuremath{\Theta \subset \Omega}}
}

\newglossaryentry{X}{
  name = $\bm{X}$ ,
  description = A random state or realisation from a sample space
}

\newglossaryentry{x}{
  name = $\bm{x}$ ,
  description = A specific state or realisation from a sample space
}

\newglossaryentry{conv_d}{
  name = $\xrightarrow[]{d}$ ,
  description = Convergence in distribution
}

\newglossaryentry{G_xi_sigma_mu}{
  name = \ensuremath{G_{\xi, \sigma, \mu}} ,
  description = {The \gls{gev} with shape parameter \ensuremath{\xi}, scale parameter \ensuremath{\sigma} and location parameter \ensuremath{\mu}}
}

\newglossaryentry{RL}{
  name = $\hat{RL}^k$ ,
  description = The estimated return level for a period with a length of $k$ blocks
}

\newglossaryentry{one}{
  name = $\one_A$ ,
  description = {The indicator function which returns a value of 1 whenever event $A$ occurs and 0 otherwise. The indicator function has the useful property \ensuremath{\Ex [\one_A] = Pr(A)}, therefore its expectation is equal to the probability of event $A$ occurring.}
}

\newglossaryentry{dcap}{
  name = $d_{cap}$ ,
  description = The power capacity rating of an asset
}

\newglossaryentry{dcrit}{
  name = $d_{crit}$ ,
  description = A critical power value above the power capacity rating of an asset
}

\begin{document}

\title{Efficient Assessment of Electricity Distribution Network Adequacy with the Cross-Entropy Method
}

\author{\IEEEauthorblockN{Julian N. Betge\IEEEauthorrefmark{1}}\thanks{\IEEEauthorrefmark{1}
The research was carried out as an intern at Alliander for an MSc thesis project. JB received funding from the German Academic Scholarship Foundation.}
\IEEEauthorblockA{\textit{Dept.~of Electrical Sustainable Energy} \\
\textit{Delft University of Technology}\\
Delft, The Netherlands \\
julian.betge@gmail.com}
\and
\IEEEauthorblockN{Barbera Droste}
\IEEEauthorblockA{\textit{IT Data \& Analytics} \\
\textit{Alliander N.V.}\\
Arnhem, The Netherlands \\
barbera.droste@alliander.com}
\and
\IEEEauthorblockN{Jacco Heres}
\IEEEauthorblockA{\textit{IT Data \& Analytics} \\
\textit{Alliander N.V.}\\
Arnhem, The Netherlands \\
jacco.heres@alliander.com}
\and[\hfill\mbox{}\par\mbox{}\hfill]
\IEEEauthorblockN{Simon H. Tindemans}
\IEEEauthorblockA{\textit{Dept.~of Electrical Sustainable Energy} \\
\textit{Delft University of Technology}\\
Delft, The Netherlands \\
s.h.tindemans@tudelft.nl}
}

\IEEEpubid{\parbox{\columnwidth}{\copyright 2021 IEEE. Personal use of this material is permitted. Permission from IEEE must be obtained for all other uses, in any current or future media, including reprinting/republishing this material for advertising or promotional purposes, creating new collective works, for resale or redistribution to servers or lists, or reuse of any copyrighted component of this work in other works.}\hspace{\columnsep}\makebox[\columnwidth]{ }}

\maketitle

\IEEEpubidadjcol

\begin{abstract}
Identifying future congestion points in electricity distribution networks is an important challenge distribution system operators face. A proven approach for addressing this challenge is to assess distribution grid adequacy using probabilistic models of future demand. However, computational cost can become a severe challenge when evaluating large probabilistic electricity demand forecasting models with long forecasting horizons. In this paper, Monte Carlo methods are developed to increase the computational efficiency of obtaining asset overload probabilities from a bottom-up stochastic demand model. Cross-entropy optimised importance sampling is contrasted with conventional Monte Carlo sampling. Benchmark results of the proposed methods suggest that the importance sampling-based methods introduced in this work are suitable for estimating rare overload probabilities for assets with a small number of customers.
\end{abstract}


\section{Introduction}

The energy transition has a considerable impact on electricity distribution networks, which increasingly accommodate distributed energy resources, such as renewable generation or storage. The challenge lies not only in the unpredictable nature of renewable generation, but also in the conventional design of distribution networks which facilitates unidirectional flows from the transmission level to end consumers. To enable adequate distribution network planning under these changing circumstances, new approaches and tools are needed. The review of suitable new approaches given \mbox{in \cite{pilo2013new}} underlines that these should be probabilistic, model risk explicitly, consider load and generation time-series, employ a multi-objective optimisation framework and incorporate a mix between network and no-network solutions. The merits of a novel planning tool realising these requirements are demonstrated e.g. in \cite{celli2013comparison}, where the benefits of integrating distributed storage over traditional grid reinforcement are assessed.

A specific challenge faced by \glspl{dso}, for which new planning approaches are needed, is to identify future congestion points in the network. This problem is a type of system adequacy assessment, where a probabilistic model of future demand is compared with the physical network capacity. Future demand may be modelled in a bottom-up manner, reflecting the uncertainty arising from various high-level technology diffusion scenarios \cite{bernards2019development}.

Focusing on this challenge and in response to the need for new network planning tools, Alliander is developing the \gls{andes} model for its subsidiary, the Dutch \gls{dso} Liander. In its current form, the load model makes use of measured load data from monitored customers, but the demand of unmonitored customers is represented by average category profiles. The use of averaged profiles is problematic, as these are smoother and less stochastic, leading to the underestimation of peaks and troughs. An extension to the \gls{andes} model which randomly assigns measured smart meter profiles to unmonitored customers was proposed in \cite{TomValckxThesis}. This was shown to improve the prediction of peak loads, but the \gls{mc} simulation approach came at a drastically increased computational cost, preventing its adoption in the regular model. 

In the context of \gls{mc} assessment of system adequacy, variance reduction techniques can be employed to speed up the convergence of risk metric estimates. Among these, \gls{is} has a particularly large variance reduction potential \cite{rubinstein2016sim_and_MC}, however it is necessary to identify a suitable biasing distribution. To accomplish the latter in an automatic manner, the Kullback–Leibler divergence is used in the \gls{ce} method. The approach has been successfully applied in power system reliability in previous studies, e.g. \cite{da_silva2010_CE_binom_distrib,tomasson2016_IS_CE}.

This paper aims to contribute to the further development of the \gls{andes} model by showing the potential of applying the \gls{ce} method in the given context. To this end, firstly an electricity demand model is specified, employing a similar \gls{mc} simulation approach as in \cite{TomValckxThesis}. Secondly, a parameterisation of the demand model is proposed which is suitable for applying the \gls{ce} method. Substantial speedups in the computation of overload probabilities can be obtained with the considered \gls{ce} approach, but its effectiveness depends on the magnitude of the quantity being estimated.

\section{Network Asset Congestion Model}
\label{sec:demand_model}

A metric of fundamental importance in capacity planning is the network asset overload probability which indicates where congestion problems are likely to arise. A variety of assets could be considered for such analysis, but in this work we focus on (transformers in) substations. The risk metric of interest here is thus the probability of the stochastic power demand $D$ (for a given asset) 
to exceed the power rating of an asset $d_{cap}$. For the remediation of congestion issues, it matters whether congestion occurs because the energy demanded by customers cannot be transported to them (risk of positive overload) or the energy generated by customers with PV systems cannot be transported away from them (risk of negative overload). The probabilities of both types of overloads are represented by the risk metrics $r_{+}$ and $r_{-}$:
    \begin{align}
        r_{+} &=  \mathbb{E}_D [\one _{D\, >\, d_{cap}}], \label{eq:rplus} \\
        r_{-} &= \mathbb{E}_D [\one _{D\, <\, -d_{cap}}], \label{eq:rminus}
    \end{align}   
    where $\one$ is the indicator function and the expectation is taken over all realisations of the power demand $D$ for the network asset of interest.

\subsection{Relation to the ANDES Model}

The \gls{andes} model relies on a bottom-up approach to obtain time-resolved asset load profiles by aggregating load profiles of individual customers. For different diffusion scenarios of key low-carbon technologies, the approach allows to predict which assets might be overloaded in the future. The model covers the entire Alliander power grid, has a time horizon of up to 40 years with quarter-hourly resolution and currently comprises five future scenarios which entails a considerable computational effort and large volumes of output data \cite{Sande2017andes}.

An essential aspect of the modelling approach presented here is to introduce stochasticity in the electricity demand behaviour of certain customers in order to obtain a probabilistic range of possible demand behaviours. In this aspect, the modelling approach differs substantially from the \gls{andes} model which currently makes deterministic predictions. A closer description of the \gls{andes} model can be found in \cite{Sande2017andes}. An in-depth comparison of the model presented here, the \gls{andes} model and the model proposed by Valckx \cite{TomValckxThesis} is available in \cite{Thesis_Julian_B}.

\subsection{Bottom-up Demand Model}

The electrical load on an asset is computed as the sum of individual demand profiles assigned to the customers connected to the asset. Only active power is considered, and losses are neglected. Each demand profile is a time series containing quarter-hourly power consumption averages, and is based on measurements from the year 2018, totalling 35,040 time steps for the entire year. The assignment of demand profiles to customers is based on the categorisation of customers used in the \gls{andes} model, a detailed break-down of which can be found in \cite{Thesis_Julian_B}. In the following, the grouping relevant for the modelling approach adopted here is presented. Three groups of demand profile types can be distinguished therein:

\subsubsection{Smart meter profiles}

A set of smart meter profiles of monitored customers of the Liander grid is used as the basis for the modelling of most categories of small unmonitored customers (\emph{Group 1}). Individually measured smart meter profiles were anonymised, but the energy-behavioural category that each profile belongs to is known. Each category contains at least 50 profiles to guarantee anonymisation. The data is owned by the \gls{dso} Liander. During model construction, filtering steps were carried out to address data quality issues, reducing the raw set of 3,773 smart meter profiles to 3,137 profiles.

Next, each energy-behavioural category was subdivided by binning the smart meter profiles and the modelled customers of each category based on their total yearly \mbox{consumption $\gamma$}. A target number of 50-100 smart meter profiles per bin was deemed appropriate leading to the formation of two to four bins per category. Quantile binning was used, so that the number of smart meter profiles per bin was approximately equal within each category. The customers to be modelled were likewise quantile binned according to their yearly consumption, forming the same number of bins per category as for the smart meter profiles. Customers are randomly assigned a profile from their associated category bin, scaled to the customer's measured yearly consumption.

While the energy-behavioural categories are those used in the \gls{andes} model, the binning scheme is a new aspect introduced in this work. Overall, group 1 comprises 13 energy-behavioural categories which are split in 2-4 bins each, leading to a total of 48 bins.

\subsubsection{Telemetry profiles}

\emph{Group 2} comprises several thousand individual larger, commercial customers whose power consumption was measured telemetrically. These telemetric measurements are directly used to model the respective customers.

\subsubsection{Average category profiles}

\emph{Group 3} contains 30 different types of profile categories: two energy-behavioural categories for which no smart meter data is available (otherwise they would be placed in group 1); 20 categories based on a classification of economic activities in which large unmonitored customers fall; and eight categories based on a classification of grid connection ratings, used for the few not otherwise categorised customers. The common trait of this diverse group is that all customers in it are modelled by scaled average category profiles derived from the described categorisations. The share of this group in overall consumption is rather low, therefore the use of average category profiles is deemed acceptable.

\subsection{Probabilistic Description of the Demand Model}

We proceed to formalise the demand model described above. In the following, random variables are capitalised and vectors are set in bold. The stochastic power demand on a given network asset for all quarter-hourly time steps of a year is obtained by summing the demand profiles of all customers connected to the asset:
\begin{equation}
    \bm{D}(\bm{\Pi}) = \sum_{i = 1}^{n_s} \gamma_i \cdot \bm{s}_{b_i, \Pi_i} + \sum_{p = 1}^{n_l} \bm{l}_{p} + \sum_{q = 1}^{n_a}  \gamma_q \cdot \bm{a}_{c_q}  ,
\label{eq:demand_model}
\end{equation}
where
\begin{itemize}
    \item $\bm{D}(\bm{\Pi})$ is a resulting random annual asset demand trace or time series
    \vspace{-6pt}
    \begin{equation*}
        \bm{D}(\bm{\Pi}) = \{{D}_t(\bm{\Pi})\} = \{{D}_1(\bm{\Pi}), \dots, {D}_{35040}(\bm{\Pi})\}  ,
    \vspace{-6pt}
    \end{equation*}
    \item $\bm{\Pi}$ is a random vector denoting a random selection of smart meter profiles,
    \item $\bm{s}_{b_i, \Pi_i}$ is a normalised smart meter profile randomly drawn from bin $b_i$ of customer $i$,
    \item $\bm{l}_p$ is the telemetry profile of customer $p$,
    \item $\bm{a}_{c_q}$ is the average category profile of category $c_q$ of customer $q$,
    \item $\gamma_i$ and $\gamma_q$ are the total yearly energy consumption of customers $i$ and $q$, respectively and
    \item $n_s$, $n_l$ and $n_a$ denote the total number of customers in groups 1-3 of the modelled asset, respectively.
\end{itemize}
\vspace{8pt}
The random profile selection vector
\begin{equation}
    \bm{\Pi} = (\Pi_1, \Pi_2, \dots, \Pi_i, \dots, \Pi_{n_s})
\end{equation}
has length $n_s$ and indexes the smart meter profiles $\bm{s}_{b_i, \Pi_i}$, randomly drawn from bin $b_i$ of each small unmonitored customer $i$ connected to the asset. The elements of $\bm{\Pi}$ are the uniform discrete random variables $\Pi_i$, which have the sample space $\Omega_{\Pi_i} = \{ 1, \dots, n_{b_i} \}$, with $n_{b_i}$ being the number of smart meter profiles in bin $b_i$ that customer $i$ belongs to. The random selection of smart meter profiles for all group 1 customers gives rise to the stochastic properties of the demand model.

\section{Monte Carlo Risk Estimation}
\label{sec:MC_methods}

Having defined a probabilistic demand model and the risk metrics $r_+$ \eqref{eq:rplus} and $r_-$ \eqref{eq:rminus}, this section details how various \gls{mc} methods can be used to estimate these risks. In the following, only the relations for the metric $r_{+}$ for positive overloads are shown and the subscript `$+$' is dropped for brevity. The relations for the metric $r_{-}$ follow by analogy.

\subsection{Conventional Monte Carlo Method}

The sample space of the demand model, which contains all possible demand states that can be assumed, has two dimensions: the assignment of smart meter profiles to customers 
and the time 
at which the selected profiles are evaluated. There is a natural hierarchy between these dimensions: the selection of random profiles occurs first and carries a larger overhead than the selection of time steps. Moreover, the profile selection space is vastly larger than the set of available time steps. 

Together, these considerations lead to a hierarchical \gls{mc} estimation scheme. The `overload fraction' of each randomly selected profile vector $\bm{\pi}_j$ and random sample $\bm{\theta}_j$ of $m$ time steps, $\{\theta_{j,1}, \dots, \theta_{j,t}, \dots, \theta_{j,m}\}$ with $\theta_{j,t} \in \{1, \dots, 35040\}$, is quantified using the impact function $H(\cdot)$ defined as:
\begin{equation}
    H(\bm{\pi}_j, \bm{\theta}_j) =  \frac{1}{m} \sum_{t = 1}^{m} \one _{D_{\theta_{j,t}}(\bm{\pi}_j)\, >\, d_{cap}} .
\label{equ:impact_funct_final}
\end{equation}
The \gls{mc} estimator for the overload probability is then the mean
\begin{equation}
    \hat{r}_{MC} = \frac{1}{n} \sum_{j = 1}^{n} H(\bm{\pi}_j, \bm{\theta}_j). \label{eq:mcregular}
\end{equation}
taken over the impact values of $n$ randomly selected profile vectors and time step samples.

As a special case, we define the \emph{reference method}, which computes entire annual asset demand traces with $m=$ 35,040 quarter-hourly time steps, i.e. 
\begin{align}
    \hat{r}_{ref} &= \frac{1}{n} \sum_{j = 1}^{n} H(\bm{\pi}_j,\bm{\theta}^* ) \label{eq:mcref} \\
    \bm{\theta}^* &= \{1,\dots,35{,}040 \}.
\end{align}
We expect the reference method to be less efficient than the generic \gls{mc} method \eqref{eq:mcregular}, because more time is spent on analysing highly dependent samples in sequential time slots. 

In both cases, the coefficient of variation or relative error of the estimator can be estimated (cf. \cite{rubinstein2016sim_and_MC}) using
\begin{equation}
    \beta_{\hat{r}} = \frac{\hat{\sigma}_H}{\hat{r} \sqrt{n} },
\label{equ:RE_ref}
\end{equation}
where $n$ denotes the number of profile selections sampled and $\hat{\sigma}_H$ is the sample standard deviation of the samples $H(\bm{\pi}_j,\bm{\theta}_j)$.

\subsection{Importance Sampling of Profile Selections}
\label{subsec:IS_of_profile_selections}

Variance reduction techniques can lead to substantial increases in the efficiency of estimating a quantity of interest. Among these, \gls{is} is the most fundamental technique \cite{rubinstein2016sim_and_MC} and lends itself to application to the problem at hand. The core idea hereby is to sample the most spiky smart meter profiles within each bin with higher probability in order to increase the frequency of overload events. 

Before being able to apply \gls{is} to the selection of profiles, the given problem needs to be parameterised appropriately. The parameterisation chosen is based on the idea to divide the smart meter profiles of each bin used in the demand model in two sets -- a set of very spiky profiles and a set of profiles with more average characteristics. 

The following approach was used to identify the most spiky profiles. Let $\bm{s}_b = \{ s_{b,t} \}$ be a smart meter profile from a given bin $b$ and $\Tilde{\bm{s}}_b = \{ \Tilde{s}_{b,t} \}$ the median profile of that bin, obtained by computing the median power consumption of all profiles in the bin per time step. For brevity, the bin indices $b$ are dropped below. The metrics $\Delta_+$ and $\Delta_-$ have been calculated for each profile in each bin by evaluating
\begin{equation}
    \Delta_+(\bm{s}) = \sum_{t = 1}^{35,040} h(t),\ h(t) = 
        \begin{cases}
            (s_t - \Tilde{s}_t)^2 & s_t > \Tilde{s}_t \\
            0 & \text{otherwise} ,
        \end{cases}
\end{equation}
and
\begin{equation}
    \Delta_-(\bm{s}) = \sum_{t = 1}^{35,040} h(t),\ h(t) =  
        \begin{cases}
            (s_t - \Tilde{s}_t)^2 & s_t < \Tilde{s}_t \\
            0 & \text{otherwise} .
        \end{cases}
\end{equation}

Using the threshold parameter $q_{spiky}$, a profile $\bm{s}$ was assigned to the set of spiky profiles for positive overloads whenever $\Delta_+(\bm{s}) \geq Q_{\Delta_+}(p = q_{spiky})$, where $Q_{\Delta_+}(p)$ is the quantile function of the median deviation metric $\Delta_+$ which returns the estimated sample $p$-quantile. The two sets formed in this way for each bin $b$ are denoted by $\mathcal{S}_{b,spiky}$ and $\mathcal{S}_{b,smooth}$. Analogously, the decision rule for negative overloads was to assign a profile $\bm{s}$ to the spiky set whenever $\Delta_-(\bm{s}) \geq Q_{\Delta_-}(p = q_{spiky})$.

The random assignment of profiles from a bin to each small, unmonitored customer (group 1) can now be parameterised in two steps. First, the customer is associated to the `spiky' set with probability $u_i$ and to the `smooth' set otherwise, where the spiky set assignment probabilities $u_i$ are determined according to
\begin{equation}
    u_i = \frac{| \mathcal{S}_{b_i,spiky} |}{| \mathcal{S}_{b_i,spiky} | + | \mathcal{S}_{b_i,smooth} |}.
\label{eq:u_i_success_initial}
\end{equation} Then, a profile is randomly chosen with uniform probability from the appropriate set ($\mathcal{S}_{b_i,spiky}$ or $\mathcal{S}_{b_i,smooth}$) of the bin $b_i$ to which the customer belongs.

The overall sampling distribution $f$ of a modelled asset, from which random profile selections $\bm{\Pi}$ are drawn, can therefore be described as a chain of $n_s$ Bernoulli trials
\begin{equation}
    f(\bm{x}; \bm{u}) = \prod_{i = 1}^{n_s} (u_i)^{x_i} \cdot (1 - u_i)^{1 - x_i} ,
\label{eq:f_u_x_orig}
\end{equation}
parametrised with the spiky set assignment probabilities $\bm{u} = (u_1, \dots, u_i, \dots, u_{n_s})$ of the $n_s$ customers modelled by a random smart meter profile (group 1).  The relation determines the probability $f(\bm{x}; \bm{u})$ of encountering a given set assignment $\bm{x} = (x_1, \dots, x_i, \dots, x_{n_s})$ of the $n_s$ customers, with $x_i \in \{0, 1\}$ and $1$ representing the spiky set.

Parameterising the demand model in the described way enables \gls{is} using a modified vector of spiky set assignment probabilities $\bm{v}$, where the probabilities of sampling spiky profiles differ from the original probabilities $\bm{u}$. To correct the purposefully introduced bias due to sampling according to the spiky set probabilities $\bm{v}$, importance weights are used:
\begin{equation}
    W(\bm{x}; \bm{u}, \bm{v}) = \frac{f(\bm{x}; \bm{u})}{g(\bm{x}; \bm{v})} = \frac{\prod_{i = 1}^{n_s} ( u_i)^{x_i} \cdot (1-u_i)^{1 - x_i}}{\prod_{i = 1}^{n_s} ( v_i)^{x_i} \cdot (1-v_i)^{1 - x_i}} ,
\label{equ:IS_weights}
\end{equation}
where $g(\bm{x}; \bm{v})$ is the \gls{is} sampling distribution. Let $\bm{x}_1, \dots, \bm{x}_n$ be an \gls{iid} sample of set assignment vectors from $g(\cdot; \bm{v})$ with the profile selections $\bm{\pi}_1(\bm{x}_1), \dots, \bm{\pi}_n(\bm{x}_n)$ sampled from the appropriate sets according to these vectors. Furthermore, let $\bm{\theta}_1, \dots, \bm{\theta}_n$ be a set of \gls{iid} samples of $m$ time steps, containing a sample of time steps for each of the $n$ profile allocations. The \gls{is} estimator is then:
\begin{equation}
    \hat{r}_{IS} = \frac{1}{n} \sum_{j = 1}^{n} H(\bm{\pi}_j(\bm{x}_j), \bm{\theta}_j) \cdot W(\bm{x}_j; \bm{u}, \bm{v}).
    \label{eq:IS_estimator_specific}
\end{equation}
The relative error is again estimated according to \eqref{equ:RE_ref}, but using the sample standard deviation of the product $H(\cdot)W(\cdot)$.

\subsection{Application of the Cross-Entropy Method}

Although \eqref{eq:IS_estimator_specific} yields a valid estimator for any (non-degenerate) choice of $\bm{v}$, its performance can vary greatly. To target overload events more precisely and thus achieve better variance reduction, the probability to sample spiky profiles can be automatically increased for those customers only which are crucially involved in causing overloads using the \gls{ce} method. 
This minimises the Kullback-Leibler divergence (a distance measure) between the \gls{is} distribution $g(\cdot; \bm{v})$ for a given asset and the theoretically optimal (but unknown) \gls{is} distribution $g^*$. If the distribution whose parameters are being \gls{ce}-optimised belongs to an exponential family, an analytical solution to the optimisation problem can be obtained, which is the case for the Bernoulli distribution used here. 
The analytical solution is then estimated using a sampling-based formula -- the \gls{ce} method thus combines optimisation and estimation \cite{rubinstein2016sim_and_MC}.

When using the \gls{ce} method in the context of rare event sampling, as done here, finding the near-optimal \gls{is} distribution is often a rare event estimation problem itself and would require large sample sizes. To circumvent this difficulty, a sequence of intermediate \gls{is} distributions can be estimated which gradually approach the near-optimal \gls{is} distribution. In this section, a sequential \gls{ce} algorithm for the estimation of overload probabilities of a distribution network asset is described, adapted from \cite{da_silva2010_CE_binom_distrib}. For a more a general discussion of the \gls{ce} method, the interested reader is referred to \cite{rubinstein2016sim_and_MC, biondini2015_rare_event_sim_IS}.

\begin{enumerate}
    \item Set the sample size for an iteration of the \gls{ce} optimisation $n_{opt}$, the multilevel parameter $\rho$ and the smoothing parameter $\alpha$ to appropriate values. Set the maximum sample size $n_{max}$, the no-overload sample size $n_{max, zero}$ and the target relative error $\beta_{target}$ used as stopping criteria. Choose a value for the threshold parameter $q_{spiky}$ and assign the smart meter profiles in each bin to the `spiky' and `smooth' sets according to the approach described in \cref{subsec:IS_of_profile_selections}.
    \item Initialise the \gls{is} distribution parameter vector $\bm{v}$ to be estimated as $\hat{\bm{v}}_0 = \bm{u}$ to start with conventional \gls{mc} sampling. Set the iteration counter of the \gls{ce} optimisation process to $k = 1$. Set the initial optimisation threshold to half of the asset's rated power capacity: $d_{opt} = 0.5 \cdot d_{cap}$.
    \item Generate an \gls{iid} sample of set assignment vectors $\bm{x}_1,\dots,\bm{x}_{ n_{opt}}$ which follow $g(\cdot; \bm{\hat{v}}_{k-1})$. Then sample profile selections $\bm{\pi}_1(\bm{x}_1), \dots, \bm{\pi}_{n_{opt}}(\bm{x}_{n_{opt}})$ accordingly. Also, generate a set of \gls{iid} samples of $m$ time steps, $\bm{\theta}_1, \dots, \bm{\theta}_{n_{opt}}$. Obtain the maximum asset load for each of the $n_{opt}$ profile selection samples and store these in the vector $\bm{l}_{max}$.
    \item Update $d_{opt}$ as the \mbox{$(1 - \rho)$-quantile} of $\bm{l}_{max}$, thus $d_{opt} = Q_{\bm{l}_{max}}(1 - \rho)$.
    \item To update the probability vector for sampling spiky profiles, evaluate for each customer $i$ the equation
    \begin{equation}
    \hat{v}_{k, i}' = \frac{\sum_{j = 1}^{n_{opt}} \tilde{H}(\bm{\pi}_j(\bm{x}_j), \bm{\theta}_j) \cdot W(\bm{x}_j; \bm{u}, \bm{\hat{v}}_{k-1}) \cdot x_{j,i}}{\sum_{j = 1}^{n_{opt}} \tilde{H}(\bm{\pi}_j(\bm{x}_j), \bm{\theta}_j) \cdot W(\bm{x}_j; \bm{u}, \bm{\hat{v}}_{k-1})} ,
    \label{eq:CE_update}
    \end{equation}
    where $\tilde{H}(\cdot)$ uses $d_{opt}$ instead of $d_{cap}$ as the capacity threshold to determine overload events. Subsequently, carry out smoothed updating by evaluating $\hat{\bm{v}}_{k} = \alpha \cdot \hat{\bm{v}}_{k}' + (1 - \alpha) \hat{\bm{v}}_{k-1}$ and bound all updated spiky set assignment probabilities between $1 - q_{spiky}$ and $0.9$.
    \item If satisfactory convergence is achieved in the optimisation stage (i.e. $\beta_{\hat{r}_{IS}} < \beta_{target}$), terminate the algorithm. 
    \item If $k\cdot n_{opt} > n_{max, zero}$ and $d_{cap}$ has not been exceeded, terminate the algorithm, as the overload probability is close to zero.
    \item If $d_{opt} > d_{cap}$, set $n=0$ and proceed with step 9. Otherwise, increment $k$ and go back to step 3.
    \item Generate an \gls{iid} sample of $50$ set assignment vectors $\bm{x}_{n+1},\dots,\bm{x}_{n+50}$ according to $g(\cdot; \bm{\hat{v}}_{k})$, and obtain the associated random profile selections $\bm{\pi}_{n+1}(\bm{x}_{n+1}), \dots, \bm{\pi}_{n+50}(\bm{x}_{n+50})$. Also generate a set of \gls{iid} samples of $m$ time steps, $\bm{\theta}_{n+1}, \dots, \bm{\theta}_{n+50}$. Set $n\leftarrow n+50$. 
    \item Estimate the overload risk using \eqref{eq:IS_estimator_specific} with $\bm{v} = \hat{\bm{v}}_k$.
    Estimate the relative error $\beta_{\hat{r}_{IS}}$ of the estimate using the weighted version of \eqref{equ:RE_ref}. If $\beta_{\hat{r}_{IS}} < \beta_{target}$ or if $k\cdot n_{opt} + n > n_{max}$, terminate the algorithm. Otherwise go back to step 9. 
\end{enumerate}

\subsection{Generalising the Importance Distribution}

If it is possible to derive a suitable general profile selection \gls{is} distribution for a wider range of assets from asset specific distributions, the optimisation stage in the \gls{ce} algorithm of the previous subsection could be skipped, potentially resulting in greater computational efficiency. For finding a generalised \gls{is} distribution, here the \gls{ce}-optimised \gls{is} distributions obtained by evaluating the \gls{ce} algorithm on 150 MV/LV substations were used. The core idea of the generalisation approach investigated here is to replace the probabilities to pick spiky profiles of \emph{individual customers} by probabilities for the 48 bins used in the demand model.

To obtain the latter, firstly the \gls{ce}-optimised $\hat{\bm{v}}_k$ of all assets with less than 80 customers were selected. Secondly, using these, the mean of all per-customer spiky set probabilities in a bin was computed yielding average bin spiky set probabilities. Finally, a thresholding was performed such that only bins with average spiky set probabilities higher than 0.15 were assigned these higher probabilities. The initial probabilities from \eqref{eq:u_i_success_initial} were assigned to the remaining bins.

\section{Results}
\label{sec:results}

The demand model and the Monte Carlo methods described in sections \ref{sec:demand_model} and \ref{sec:MC_methods} were implemented in \textit{R} version 3.6.1. 
The code was developed and run on a shared RStudio Server environment. For speed comparisons, a generic computation was carried out repeatedly to verify that computational speeds due to different workloads of the shared RStudio Server environment varied in an acceptably small range.

The convergence target for all Monte Carlo methods was set to $\beta_{target} = 0.1$, with $n_{max} =$ 20,000 and $n_{max,zero}$=10,000 as maximum sample sizes to keep computation times in reasonable limits. To determine suitable values of the remaining method parameters, the sensitivity of computation time to the variation of individual parameters in a discrete range around an experience-based default value were explored for ten representative substations. Based on these initial experiments, the number of time steps sampled per trace was set to $m =$ 2,000 for all methods and the additional parameters of the \gls{ce} algorithm were set to $n_{opt} = 500$, $\rho = 0.05$, $\alpha = 0.6$, $q_{spiky} = 0.95$ (see \cite{Thesis_Julian_B} for a more detailed description).

To benchmark the computational efficiency of the Monte Carlo methods, the abovementioned 150 MV/LV substations were employed. For the reference method (abbr. as \textit{Ref}), the conventional \gls{mc} sampling (abbr. as \textit{MC}) and \gls{ce}-optimised profile selection \gls{is} (abbr. as \textit{CE-IS}) nine replicates were computed for each substation. Due to limited computational resources, for the generalised bin spiky set probability \gls{is} method (abbr. as \textit{Gen-IS}) five replicates were computed.

\subsection{Accuracy of the Investigated IS-MC Methods}

From theoretical considerations it follows that both conventional \gls{mc} sampling and \gls{is}-\gls{mc} yield unbiased estimates. However, contrasting the \gls{is}-\gls{mc} benchmarking results with conventional \gls{mc} sampling results showed that inaccurate risk estimates were often obtained for substations with a large number of customers.

The number of substation customers determines the number of parameters to be estimated and thus the dimensionality of the problem. 
It is a known issue that in high-dimensional settings the likelihood ratio (given in \eqref{equ:IS_weights}) degeneracy problem leads to unstable \gls{is} distributions \cite{chan2012improved}. Concretely the issue arises e.g. when more than 15-20 customers are assigned high spiky set probabilities by the \gls{ce} algorithm in a suboptimal way, which results in nearly all samples having extremely small importance weights (see \eqref{eq:IS_estimator_specific}).

Additional experiments for a single substation (results not shown) indicate that the likelihood degeneracy problem can be ameliorated when the optimisation sample size is increased. However, for the investigated case, $n_{opt} \geq$ 20,000 was necessary to obtain an estimate in the correct order of magnitude which strongly reduces computational efficiency.

Therefore, for the speedup quantification in the next subsection, a filtering step was included: results of both \gls{is} methods were compared to results of the reference method using Welch's $t$-test, and omitted if estimates were significantly different ($\alpha=0.05$). In practice, this mostly excluded assets with more than 80 customers. Future research could investigate limiting the dimensionality of the model parameterisation or implement the improved \gls{ce} method proposed in \cite{chan2012improved}.

\subsection{Efficiency of the Investigated MC Methods}

The computation times $t$ until reaching $\beta_{target} = 0.1$ were measured for the computational efficiency benchmarking of the three considered \gls{mc} sampling methods against the reference method. 
For those cases where the estimate did not converge to $\beta_{target} = 0.1$ before reaching $n_{max} =$ 20,000, an extrapolation based on the expression for the relative error from \eqref{equ:RE_ref} was made to approximate the total computation time needed for convergence.

\begin{figure}[htbp]
\centerline{\includegraphics[width=0.488\textwidth]{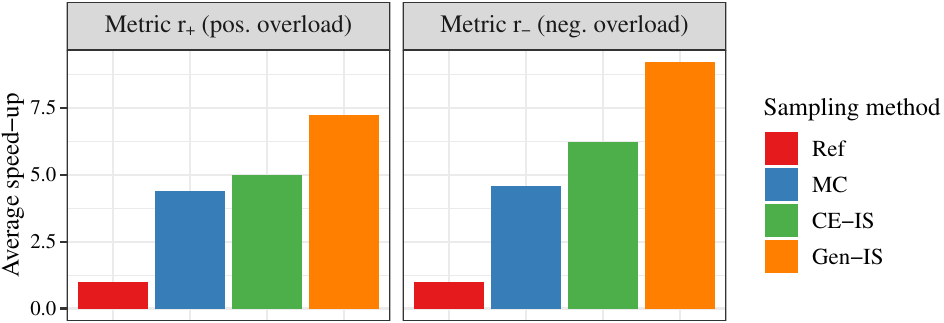}}
\caption{Comparison of the average speed-up $\bar{t}_{ref}/t_{method}$ for all methods with respect to the reference method.}
\label{fig:avg_speedup}
\end{figure}

The average speed-ups of the sampling methods with respect to the reference method in obtaining $r_+$ and $r_-$ are shown in Fig.~\ref{fig:avg_speedup}. Only nonzero estimates and cases where estimates were found to be accurate are considered here and for the remainder of this section. While all methods perform better than the reference method, it is surprising that the \textit{CE-IS} method has a similar computational speed as the \textit{MC} method. The \textit{Gen-IS} method, in turn, outperforms both, as it can reap the benefits of profile selection \gls{is} without the computational cost of the \gls{ce} optimisation required in the \textit{CE-IS} method.

\begin{figure}[b]
\centerline{\includegraphics[width=0.488\textwidth]{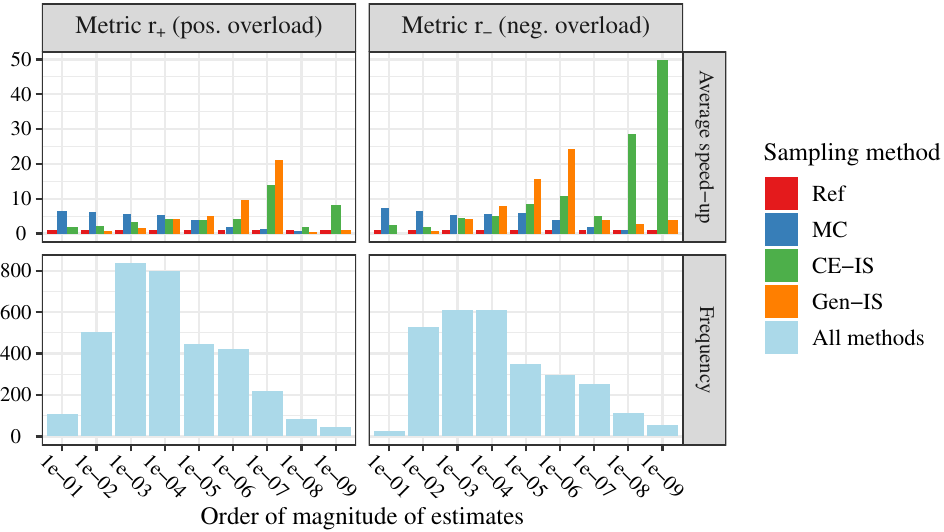}}
  \caption{Comparison of the average speed-ups $\bar{t}_{ref}/t_{method}$ (top panels) of estimates per bin for all methods with respect to the reference method. The bottom panels show histograms indicating how many estimates fall in which order of magnitude.}
  \label{fig:speed-up_per_bin}
\end{figure}

However, average speed-up comparisons across many assets with different properties hide much of the underlying detail. Generally, one of the most important determinants of the required computational effort of Monte Carlo methods is the magnitude of the quantity being estimated. Therefore, in Fig.~\ref{fig:speed-up_per_bin}, average speed-ups per order of magnitude of the estimates of metrics $r_{+}$ and $r_{-}$ are shown for all methods.

Fig.~\ref{fig:speed-up_per_bin} reveals that the profile selection \gls{is} methods were able to achieve high speed-ups between 10-50 times for estimates in the orders $10^{-5}$ and smaller in several cases, clearly outperforming conventional \gls{mc} sampling. Furthermore, the \textit{CE-IS} method shows much higher speed-ups in the orders $10^{-8}$ and $10^{-9}$ for the $r_-$ metric than the \textit{Gen-IS} method. This may indicate that the \gls{ce} optimisation of the \gls{is} distribution for each asset individually leads to well targeted \gls{is} distributions for very rare events -- which outweighs the advantage of omitting the optimisation stage in the \textit{Gen-IS} method.
Finally, the histograms in the bottom panels of Fig.~\ref{fig:speed-up_per_bin} show that most estimates fall in the orders $10^{-5}$ and larger, which are the orders of practical relevance for \glspl{dso}. This explains why the high speed-ups of the profile selection \gls{is} methods for rare events do not appear as pronounced in the average speed-ups of Fig.~\ref{fig:avg_speedup} and it suggests that conventional \gls{mc} sampling is already well suited for many practically relevant cases. A refined algorithm could adapt its sampling strategy depending on the estimate order of magnitude such that the respectively most advantageous method is employed.

\section{Conclusion}

In this paper, firstly an approach to account for the stochasticity of customer behaviour in distribution network demand modelling was presented. Secondly, Monte Carlo methods for the computationally efficient estimation of network asset overload probabilities were developed. The evaluation of the methods on 150 MV/LV substations showed that cross-entropy optimised importance sampling in the current setting works well for assets with less than 50-80 customers and shows the greatest performance advantages over conventional Monte Carlo sampling for rare event probability estimates.

\bibliographystyle{IEEEtran.bst}

\end{document}